\documentclass[12pt]{article}
\usepackage{graphicx}
\usepackage{float}
\usepackage[maxfloats=500]{morefloats}
\begin{document}

\begin{center}
{\bf{\Large  Regular Interior  Solutions to the Solution  of  Kerr which Satisfy the Weak and the Strong Energy Conditions }}
\vspace{1cm}

E. Kyriakopoulos\\
Department of Physics\\ 
National Technical University
15780 Zografou, Athens, GREECE\\
E-mail: kyriakop@central.ntua.gr\\
\end{center}
 
\begin {abstract}

The line element of a class of solutions which match to the solution of Kerr on an oblate spheroid if the two functions $ F(r)$ and $H(r)$ on which it
depends satisfy certain matching conditions is presented. The non vanishing components of the Ricci tensor $R_{\mu\nu}$, the Ricci scalar $ R$, the second order
curvature invariant $K$, the eigenvalues of the Ricci tensor, the energy density $\mu$, the tangential pressure  $P_{\perp}$, and the quantity $\mu+P_{\perp}$ are
calculated. A function $F(r)$ is given for which $R$ and $K$ and therefore the solutions are regular. The function $ H(r) $ should be such that  the solution 
it gives satisfies at least the Weak Energy Conditions (WEC). Several $H(r)$ are given explicitly for which the resulting 
solutions satisfy the WEC and also the Strong Energy Conditions (SEC) and the graphs of their $\mu$, $P_{\perp}$ and $\mu+P_{\perp}$ for certain values of their
 parameters are presented. It is shown that all solutions of the class are anisotropic fluid solutions and that there are no perfect fluid solutions in the class.

 \end{abstract}


\section{Introduction}

Soon after the discovery of the solution of Kerr \cite{Ke1} the problem of finding interior solutions to this solution became a major problem of the 
general theory of relativity. The  efforts before 1978 are described in Ref \cite{Kr1}. According to this reference since all efforts were 
unsuccessful there appear an opinion, without however some proof, that the metric of Kerr may have no other source than a black hole. Later attempts 
are described in references \cite{De1}-\cite{Ky1} and most resent attempts in references \cite{He2}-\cite{Ky2}.

To find regular interior solutions to the solution of Kerr, which satisfy at least the WEC, we consider the line element of Eq (\ref{2-1}) which 
depends on two functions $F(r)$ and $H(r) $ and which for $F(r)=r^2$ and $H(r)=-2 M r$ differs from the lime element of the solution of Kerr in 
Boyer-Lindquist coordinates \cite{Bo1} in the coefficient of $dr^2$, since this coefficient has an extra factor $cos^2(1-r/k)$. The exemption is 
necessary because it was found recently that for the Kerr like black hole spacetimes  the regularity is linked to a 
violation of the WEC around the core of the rotating black hole \cite{To1}. 

Starting from the line element of Eq (\ref{2-1}) we calculate the non vanishing components of the Ricci tensor $R_{\mu\nu}$, the Ricci scalar $R$, 
the second order curvature invariant $K$, the eigenvalues of the Ricci tensor, the eigenvalue  which corresponds to the timelike eigenvector 
of the Ricci tensor, the energy density $\mu$, the radial pressure $P_{r}$  and the tangential pressure $P_{\perp}$. For all solutions of the model we have 
the equation of state $\mu+P_{r}=0$. We calculate the normalized eigenvectors of the Ricci tensor and we prove that all solutions
of the model are for arbitrary $F(r)$ and $H(r)$ anisotropic fluid solutions. Also we prove that perfect fluid solutions do not exist in the solutions 
of our model. 

In a previous work \cite{Ky2} we presented many functions which at the matching surface $r=k$ satisfy the matching conditions of Eq (\ref{2-4}). We take
one of these functions as function $H(r)$ if it creates a solution which for some values of its parameters satisfies at least the WEC. Also we find
with numerical computer calculations the range of values of its parameters for which this happens.

Having this in mind we find five solutions which satisfy the WEC and the SEC but not the Dominant Energy Conditions (DEC).Also we draw
the graph of $\mu$ of $P_{\perp}$ and of $\mu+P_{\perp}$ of these solutions for certain values of their parameters.   
   
\section{The General Case}
Consider  the line element
\begin {eqnarray}
\label {test}
&& d^2 s = -(1 + \frac {H (r)} {F (r) + 
         a^2 cos^2\theta}) d t^2 + \frac {2 a sin^2\theta H (r)} {F (r) + a^2 cos^2\theta} dtd\phi + \nonumber \\
&& \frac {cos^2(1-r/k)(F (r) + a^2 cos^2\theta)} {F (r) + H (r) + 
     a^2} d r^2 + (F(r)+a^2cos^2\theta)d \theta^2  + \nonumber \\
&& sin^2\theta (F (r) + 
    a^2  - \frac { a^2 sin^2\theta H (r)} {F (r) + a^2 cos^2\theta}) d \phi^2
\label {2-1}
\end {eqnarray}
where
\begin{equation}
F(r)=(1-sin(1-r/k))^2k^2
\label{2-2}
\end{equation}
with $k$ a constant. This line element can be the line element of an interior solution to the solution of Kerr if certain matching conditions are satisfied.
The Darmoise matching conditions on a surface $r=k$ are continuity of the first fundamental form and continuity of the extrinsic curvature 
(second fundamental form) on this surface. With interior solution the solution with the above line element and exterior solution the solution of Kerr
since the function $F(r)$ satisfy the relations 
\begin{equation}
F(k)=k^2\>\>\>\mbox{and}\>\>\>F'(k)=2k
\label{2-3}
\end{equation}
continuity of the first and the second fundamental forms with Kerr's exterior solution in Boyer-Lindquist coordinates is obtained if the function 
$H(r)$ satisfies the relations
\begin{equation}
H(k)=-2 M k\>\>\>\mbox{and}\>\>\>H'(k)=-2 M
\label{2-4}
\end{equation}
Also we find that if relations (\ref{2-4}) are satisfied the interior metric (\ref{2-1}) and the exterior Kerr metric as well as their derivatives in
coordinates are continuous at the matching surface $ r=k $, which means that we have matching according to the matching conditions of Lichnerowicz. 
The coordinates used are admissible.
In the whole paper prime means derivative with respect to $ r $. 
In Boyer-Lindquist coordinates the matching surface $r=k$ 
is an oblate spheroid \cite{Gu1}. 

We shall proceed in the calculations using the relation 
\begin{equation}
\cos ^2 \left(1-\frac{r}{k}\right)=\frac{F'(r)^2}{4 F(r)}
\label{2-5}
\end{equation}
to eliminate the  $\cos ^2 (1-\frac{r}{k})$ from the line element of relation (\ref{2-1}). 
Then the non-vanishing components of the Ricci tensor 
$ R_{\mu\nu} $   the Ricci scalar $ R $ and
 the second order curvature invariant $ K =R_{\mu\nu\rho\sigma}R^{\mu\nu\rho\sigma} $ (Kretschmann scalar) as functions of  $ F(r), H(r) $ and  $ x $ defined by the relation
\begin{equation}
x=cos\theta
\label{2-6}
\end{equation}
are 
\begin {eqnarray}
\label {test}
&&R_{tt}= \frac {1} {F' (r)^3\left (a^2 x^2 + 
       F (r) \right)^3} (2 F (r) (a^2 + 
     F (r)) F' (r) H'' (r) (a^2 x^2 + F (r))   \nonumber \\
  &&  + H (r) (-2 F (r) F'' (r) H' (r) (a^2 x^2 + F (r)) + 
        2 F (r) F' (r) H'' (r) (a^2 x^2 +\nonumber \\
   &&     F (r)) +  F' (r)^2 H' (r) (a^2 x^2 - F (r)) + 
        F' (r)^3 (F (r) - a^2 (x^2 - 2))) +H' (r) \nonumber \\
 &&   (F' (r)^2 (a^4 x^2 + 3 a^2 (x^2 - 1) F (r) - 
           F (r)^2) - 2 F (r) (a^2 + F (r)) F'' (r) (a^2 x^2  \nonumber \\
  &&      + F (r))) +  H (r)^2 F' (r)^3)
\label {2-7}
\end {eqnarray}
\begin {eqnarray}
\label {test}
&& R_{t\phi}=\frac {1} {F' (r)^3\left (a^2 x^2 + 
       F (r) \right)^3} (2 F (r) (a^2 + 
       F (r)) F' (r) H'' (r) (a^2 x^2 + F (r))  \nonumber \\
 &&  + H (r) (-2 F (r) F'' (r) H' (r) (a^2 x^2 + F (r)) + 
        2 F (r) F' (r) H'' (r) (a^2 x^2 + \nonumber \\
  &&      F (r)) +  F' (r)^2 H' (r) (a^2 x^2 - F (r)) + 
       F' (r)^3 (F (r) - a^2 (x^2 - 2))) +  H' (r)\nonumber \\
  &&  (F' (r)^2 (a^4 x^2 + 3 a^2 (x^2 - 1) F (r) -
  F (r)^2) -  2 F (r) (a^2 + F (r)) F'' (r) (a^2 x^2 \nonumber \\
 &&    + F (r))) + H (r)^2 F' (r)^3)
      \label {2-8}
     \end {eqnarray}
 \begin {eqnarray}
\label {test}
&& R_{rr}=\frac {1} {4 F (r) F' (r) (a^2 + F (r) + 
       H (r) ) (a^2 x^2 + F (r))}( -2 F (r) F' (r) H'' (r)\nonumber \\
  &&     (a^2 x^2  + F (r)) + 
  H' (r) (2 F (r) F'' (r) (a^2 x^2 + F (r) ) + 
     F' (r)^2 (F (r) - a^2 x^2 ))  \nonumber \\
  &&  - H (r) F' (r)^3)
\label {2-9}
\end {eqnarray}
\begin {equation}
R_{\theta\theta}=\frac {H (r) F' (r) - 
   2 F (r) H' (r)} {F' (r)\left (a^2 x^2 + F (r) \right)}
\label {2-10}
\end {equation}     
\begin {eqnarray}
\label {test}
&& R_{\phi\phi}=\frac {1} {F' (r)^3\left (a^2 x^2 + F (r) \right)^3} (x^2 - 
    1) (a^2\ (x^2 - 1) H (r)^2 F' (r)^3 + H (r)  \nonumber \\
 &&   (-2 a^2 (x^2 - 1) F (r) F'' (r) H' (r) (a^2 x^2 + F (r)) + 2 a^2
         (x^2 - 1) F (r) F' (r)  \nonumber \\
  &&       H'' (r) (a^2 x^2 + F (r)) + a^2
         (x^2 - 1) F' (r)^2 H' (r) (a^2 x^2 - F (r)) + (a^2 +   \nonumber \\
    &&     F (r)) F' (r)^3 (a^2 (x^2 - 2) - F (r))) + (a^2 + F (r))
     (2 a^2 (x^2 - 1) F (r) F' (r)  \nonumber \\
  &&   H'' (r) (a^2 x^2 + F (r)) +  H' (r) (F' (r)^2 (a^4 x^2 (x^2 - 1) - a^2 (x^2 - 3) F (r) + \nonumber \\
     &&        2 F (r)^2) -  2 a^2 (x^2 - 1) F (r) F'' (r) (a^2 x^2 + F (r)))))
\label {2-11}
\end {eqnarray}
\begin {equation}
R = \frac {-4 F (r) F' (r) H'' (r) - 
    2 H' (r)\left (F' (r)^2 - 
       2 F (r) F'' (r) \right)} {F' (r)^3\left (a^2 x^2 + 
      F (r) \right)}
\label {2-12}
\end {equation}
\begin {eqnarray}
\label {test}
&& K = \frac {1} {F' (r)^6\left (a^2 x^2 + 
       F (r) \right)^6} 4 (2 H (r)^2 F' (r)^6 (7 a^4 x^4 - 
        34 a^2 x^2 F (r) +\nonumber \\
    &&    7 F (r)^2) +  2 H (r) F' (r)^3  (a^2 x^2 + 
        F (r)) (H' (r) (2 F (r) F'' (r) (a^2 x^2 - 3 F (r))\nonumber \\
    &&    (a^2 x^2 + F (r)) + 
           F' (r)^2 (-a^4 x^4 + 34 a^2 x^2 F (r) - 13 F (r)^2)) -   2 F (r) F' (r)\nonumber \\
      &&  H'' (r) (a^2 x^2 - 3 F (r)) (a^2 x^2 + 
           F (r))) + (a^2 x^2 +   F (r))^2 (4 F (r)^2 F' (r)^2 \nonumber \\
      &&   H'' (r)^2 (a^2 x^2 + F (r))^2 - 
        4 F (r) F' (r) H' (r) H'' (r) (a^2 x^2 +   F (r)) (2 F (r)  \nonumber \\
       &&    F'' (r) (a^2 x^2 + F (r)) + 
           F' (r)^2 (3 F (r) - a^2 x^2)) +  H' (r)^2 (4 F (r)^2 F'' (r)^2  \nonumber \\
     &&  (a^2 x^2 + F (r))^2 + 
           4 F (r) F' (r)^2 F'' (r) (a^2 x^2 + F (r)) (3 F (r) -  a^2 x^2) +  \nonumber \\
        &&  F' (r)^4 (a^4 x^4 - 18 a^2 x^2 F (r) + 13 F (r)^2))))
\label {2-13}
\end {eqnarray}
The calculations of the above quantities and all calculations of the paper were done with a computer  program of Bonanos \cite{Bs1}

From Eq  (\ref{2-2}) we find that $F(r)$ and $F'(r)$ do not vanish in the interior region  $0 \leq r \leq k $ and therefore
the invariants $R$ and $K$  of Eqs (\ref{2-12}) and (\ref{2-13}) are not singular in this region if $H(r)$, $H'(r)$ and $H''(r)$ are not singular
in this region. Therefore the solution is regular in the interior region if  $H(r)$, $H'(r)$ and $H''(r)$ are not singular
in this region.

The eigenvalues $B_{1}$ and $B_{2}$ of the Ricci tensor $R_{\mu}^{\nu}$ are the following:
\begin {eqnarray}
\label {test}
&& B_{1}= B_{r}=\frac {1} {F' (r)^3\left (a^2 x^2 +  F (r) \right)^2}  
       (2 F (r) F'' (r) H' (r)\left (a^2 x^2 +  F (r) \right)  \nonumber \\
  &&     - 2 F (r) F' (r) H'' (r)\left (a^2 x^2  +F (r) \right) + 
    F' (r)^2 H' (r)\left (F (r) - a^2 x^2 \right) - H (r)  \nonumber \\
 &&   F' (r)^3)
\label {2-14}
\end {eqnarray}
\begin{equation}
B_{2}=B_{\theta}=\frac{H(r) F'(r)-2 F(r) H'(r)}{F'(r) \left(a^2 x^2+F(r)\right)^2}
\label{2-15}
\end{equation}

To find the  eigenvalues $ B_{t} $ and $ B_{\phi}$ of the timelike eigenvector 
$ (u_{t})^{\mu }$ and the spacelike eigenvector $ (u_{\phi})^{\mu }$  of  the  Ricci tensor  $ R_{\mu}^{\nu} $ respectively
we consider  the  eigenvalue equation  $ R_\mu^\nu (u_i)^\mu=B_i(u_i)^\nu  $.
  Writing for  the eigenvectors $ (u_i)^\mu $ ,  $ i=t $ and $ \phi $   which correspond to the eigenvalues $ B_t $ and $ B_\phi $ 
\[(u_i)^\mu = \left(\begin{array}{c}
                     b_i\\0\\0\\c_i
                    \label{3-1-13'}
                    \end{array}\right)
\]
we  get  the  relations 
\begin{equation}
(R_t^t-B_i) b_i+R_t^\phi c_i=0\>\>\>and\>\>\>
R_\phi^t b_i+(R_\phi^\phi-B_i) c_i=0
\label{2-16}
\end{equation}
Therefore we have 

\begin{equation}
\frac{c_i}{b_i}=-\frac{R_t^t-B_i}{R_t^\phi}=-\frac{R_\phi^t}{R_\phi^\phi-B_i}
\label{2-17}
\end{equation}
Then we get  $ (u_i)_\mu (u_i)^\mu = b_i^2 V_i(r,x) $,  where 

\begin{equation}
V_i(r,x)=g_{tt}- 2\frac{R_\phi^t}{R_\phi^\phi-B_i}g_{t \phi} + (\frac{R_\phi^t}{R_\phi^\phi-B_i})^2 g_{\phi\phi}
\label{2-18}
\end{equation}
If  $V_i(r,x )$ is negative (positive) in the interior region $ 0\leq r \leq k$  the eigenvalue $ B_i $ is eigenvalue of 
a timelike  ( spacelike ) eigenvector  $ (u_i)^\mu $. For the metric of Eq (\ref{2-1}) with the $cos^2(1-r/k)$ 
replaced as in Eq (\ref{2-5}) and arbitrary $ F(r)$ and $H(r)$  we get 
 
\begin{equation}
V_1(r,x)=-\frac {\left (a^2 + F (r) + H (r) \right)\left (a^2 x^2 + 
     F (r) \right)} {\left (a^2 + F (r) \right)^2}
\label{2-19}
\end{equation}

\begin{equation}
V_2(r,x)=\frac{F(r)+a^2 x^2}{a^2(1-x^2)}
\label{2-20}
\end{equation}
For the $F(r)$ of Eq (\ref{2-2}) we have $ V_2(r,x) > 0$ and we shall choose $a$ such that   
\begin{equation}
a^2 + F (r) + H (r)> 0
\label{2-20a}
\end{equation}
in which case we have $ V_1(r,x)< 0 $. Therefore we get     
\begin{equation}
B_{t}=B_{1}=B_{r}
\label{2-20b}
\end{equation}
and     
\begin{equation}
B_{\phi}=B_{2}=B_{\theta}
\label{2-20c}
\end{equation}
From relation (\ref{2-20a}) for $r=k$ and Eqs (\ref{2-3}a) and (\ref{2-4}a) we get      
\begin{equation}
a^2+k^2-2 M k > 0
\label{2-20d}
\end{equation}     
which means that  $ k > M+(M^2-a^2)^{1/2}$
or $ k < M-(M^2-a^2)^{1/2}$. Therefore the matching occurs outside the outer horizon  or inside 
the inner horizon of the exterior solution of Kerr.
     
The energy  density $ \mu $, the radial pressure $ P_{r} $, the tangential  pressure  $ P_{\perp}=P_{\theta}=P_{\phi} $ and  the 
 quantities $ \mu+P_{r} $ 
and  $ \mu+P_{\perp} $  are given by the expressions: 
\begin {equation}
\mu=\frac {\frac {R} {2} - B_{t}} {8\pi} = \frac {k H' (r)\left 
 (\tan\left (1 - \frac {r} {k} \right) - \sec\left (1 - \frac 
{r} {k} \right) \right) + 
    H (r)} {8\pi\left (a^2 x^2 + 
       k^2\sin^2\left (1 - \frac {r} {k} \right) - 
       2 k^2\sin\left (1 - \frac {r} {k} \right) + k^2 \right)^2}
\label {2-21}
\end {equation}
 \begin {equation}
P_{r}=\frac { B_{r}-\frac {R} {2}} {8\pi} =- \frac {k H' (r)\left 
( \tan\left (1 - \frac {r} {k} \right) - \sec\left (1 - \frac 
{r} {k} \right) \right) + 
    H (r)} {8\pi\left (a^2 x^2 + 
       k^2\sin^2\left (1 - \frac {r} {k} \right) - 
       2 k^2\sin\left (1 - \frac {r} {k} \right) + k^2 \right)^2}
\label {2-22}
\end {equation}
\begin {eqnarray}
\label {test}
&& P_{\perp} = \frac {B_{\phi} - \frac {R} {2}} {8\pi} = \frac {1} {16\pi k \
(a^2 x^2 + 
        k^2 (\sin (1 - \frac {r} {k} ) - 
            2 )\sin (1 - \frac {r} {k} ) +  k^2 )^2} \nonumber \\
&& (2 k H(r)-\frac{1}{4} \sec ^3(1-\frac{r}{k})(2 k H''(r) \cos
 (1-\frac{r}{k}) (-2 a^2 x^2+k^2 (4 \sin
   (1-\frac{r}{k})  \nonumber \\
 &&  +\cos (2-\frac{2 r}{k}))-3 k^2)+H'(r) (2 (2 a^2 x^2+k^2) \sin
   (1-\frac{r}{k})+2 k^2 (4- \nonumber \\
 &&  (3 \sin (1-\frac{r}{k})) \cos (2-\frac{2 r}{k}))))
\label {2-23}
\end {eqnarray}
\begin{equation}
\mu+P_{r}=0
\label{2-24}
\end{equation}
\begin {eqnarray}
\label {test}
&& \mu+P_{\perp} = \frac {1} {64\pi k\left (a^2 x^2 + 
       k^2\left (\sin\left (1 - \frac {r} {k} \right) - 
           2 \right)\sin\left (1 - \frac {r} {k} \right) +  k^2 \right)^2} \nonumber \\
&& (-\sec^3 (1 - \frac {r} {k}) (2 k H'' (r)\cos (1 - \frac {r} {k}) (-2 a^2 x^2 + 
        k^2 (4\sin (1 - \frac {r} {k}) + \nonumber \\
 &&       \cos (2 - \frac {2 r} {k})) - 3 k^2) + 
     H' (r) ((4 a^2 x^2 + 3 k^2)\sin (1 - \frac {r} {k}) + 
        k^2 (-5\sin (3 -  \nonumber \\
  &&      \frac {3 r} {k}) + 
           12\cos (2 - \frac {2 r} {k}) + 4)) - 
     16 k H (r)\cos^3 (1 - \frac {r} {k})))
\label {2-25}
\end {eqnarray}
 
From Eqs (\ref{2-4}), (\ref{2-21}) and (\ref{2-22}) we find that at the matching surface $r=k$  for all $H(r)$  we get  $ \mu =P_{r}=0 $. This is a very important 
feature of our solutions.  

Expression (\ref{2-24}), which holds for all solutions of the paper, is the equation of state of these solutions. An equation of state of the form $\mu+P_r=0$
was introduced originally by Sakharov as an equation of state of a superdense  fluid \cite{Sa1}.
Gliner \cite{Gl1}  argues that the meaning of a negative pressure is that
the internal volume forces in the matter are not forces of repulsion  but forces of attraction  and also that  an object with this equation 
of state might be formed in gravitational collapse.   This equation  arises in Grand Unified Theories at very high densities 
and it is used in the cosmological inflationary senario \cite{Il1}. Also it is the equation of state in the de Sitter interior
of the gravastar  (gravitational vacuum star)  model \cite{Ma3} \cite{Vi1}. 
   
We can show that the energy-momentum tensor $ T_{\mu\nu} $ of the solutions  with the line element of Eqs (\ref{2-1}) and the $cos^2(1-r/k)$ 
replaced as in Eq (\ref{2-5}) has 
for arbitrary  $ F(r) $ and $H(r)$ the form of the energy-momentum tensor of an anisotropic fluid  solution. To do that we consider the normalized
eigenvectors $(u_{\mu})_{\nu}$   $ \mu,\nu=t,r,\theta,\phi$  of the Ricci tensor $ R_{\mu}^{\nu}$, which are given by the relations 

\begin {equation}
(u_{t})_{\mu }=\frac {\sqrt {a^2 + F (r) + H (r)}} {\sqrt {a^2 x^2 +  F (r)}} 
(-\delta_{\mu t} +  a(1 - x^2)\delta_{\mu\phi})
\label {2-26}
\end {equation}

\begin{equation}
(u_{r})_{\mu}=\sqrt{\frac{(a^2 x^2+F(r)){F'(r)}^2}{4F(r)(a^2+H(r)+F(r))}}\delta_{r\mu}
\label{2-27}
\end{equation}

\begin{equation}
(u_{\theta})_{\mu}= \sqrt{a^2 x^2+F(r)}\delta _{\theta \mu }
\label{2-28}
\end{equation}

\begin{equation}
(u_{\phi})_{\mu}= \sqrt{\frac{1-x^2}{a^2 x^2+F(r)}}(-a\delta_{t\mu}+(F(r)+a^2)\delta_{\phi\mu})
\label{2-29}
\end{equation}
From the expression for $ g_{\mu\nu}$ which is obtained from relations (\ref{2-1}) and (\ref{2-5}) and the expressions for $ R_{\mu\nu},\>R,\>\mu,\>P_{r},
\>P_{\perp},\>\>(u_{t})_{\mu} $ and  $ (u_{r})_{\mu} $  given by 
Eqs   (\ref{2-7})- (\ref{2-12}), (\ref{2-21}) - (\ref{2-23}),  (\ref{2-26}) and (\ref{2-27}) we find 
that  the  energy-momentum tensor  $ T_{\mu\nu}$  of  all solutions obtained for arbitrary  $ F(r) $  and $ H(r) $ is  given by the relation
\begin{equation}
T_{\mu\nu}=\frac{1}{8\pi}(R_{\mu_\nu}-\frac{R}{2}g_{\mu\nu})=(\mu+P_{\bot})(u_{t })_{\mu}(u_{t })_{\nu}+P_{\bot}  g_{\mu\nu}+(P_{r}-P_{\bot})(u_{r })_{\mu}(u_{r })_{\nu}
\label{2-1-30}
\end{equation}
 which is the energy-momentum tensor of an anisotropic fluid  solution \cite{He1}.  

We shall examine if functions $H(r)$ exist for which the line element (\ref{2-1}) is the line
element of a perfect fluid solution. To do that we shall use the relation $B_{r}=B_{\theta}$ which holds for perfect fluid solutions. From this relation and Eqs 
(\ref{2-14}) and (\ref{2-15}) we get      
\begin{eqnarray}
\label{test}
 && 2 F(r) F''(r) H'(r) \left(a^2 x^2+F(r)\right)-2 F(r) F'(r) H''(r) \left(a^2 x^2+F(r)\right)+ \nonumber \\
 && F'(r)^2 H'(r) \left(3 F(r)-a^2 x^2\right)-2 H(r) F'(r)^3=0
\label{2-31}
\end{eqnarray}
from which we get from the vanishing of the coefficients of the powers of $x$    
\begin {equation}
F (r)(2 F (r) F'' (r) H' (r) - 2 F (r) F' (r) H'' (r) + 
    3 F' (r)^2 H' (r)) - 2 H (r) F' (r)^3=0
\label {2-32}
\end {equation}
\begin {equation}
-2 F (r) F'' (r) H' (r) + 2 F (r) F' (r) H'' (r) + 
  F' (r)^2 H' (r) = 0
\label {2-33}
\end {equation}
Substituting in Eq (\ref{2-33}) the $ F(r)$ of expression (\ref{2-2}), solving the resulting differential equation, and imposing on the solution 
$H(r)$ the matching conditions (\ref{2-4}) we get  
\begin {equation}
H(r)=-2 k M\left (1 - \sin\left (1 - \frac {r} {k} \right) \right)=-2 k M (F(r))^{1/2}
\label {2-34'}
\end {equation}
Since the expressions (\ref{2-2}) and (\ref{2-34'}) of $F(r)$ and $H(r)$ respectively satisfy Eq (\ref{2-32}) also, these expressions are the solution 
of the system of Eqs (\ref{2-32}) and (\ref{2-33}) which satisfies the matching conditions. But the solution with the line element of Eq (\ref{2-1})
and the $F(r)$ and $H(r)$ of Eqs (\ref{2-2}) and (\ref{2-34'}) respectively is the solution of Kerr.Therefore perfect fluid solutions we try to find do not exist.

All solutions we shall consider explicitly have the $ F(r)$ of relation (\ref{2-2}). To complete the solutions we must find the functions
 $H(r)$. These functions besides the relations (\ref{2-4}) which they should satisfy they should give interior solutions which satisfy at least
  the WEC. In fact in the following Examples we shall find interior solutions whose $H(r)$ depend on a parameter $b$ and we shall 
  determined the values of $b$ for which the resulting solutions satisfy the WEC and the SEC. The WEC are defined by the relations 
  $\mu \geq0 $ and $\mu+P_{i} \geq 0 $,  $i=r, \theta, \phi$,  the SEC by the relations $ \mu+P_{i}\geq0 $ and $\mu+P_{r} +P_{\theta}+ 
  P_{\phi}\geq 0 $ 
 and the DEC by the relations $\mu\geq 0$ and $ -\mu\leq P_{i} \leq \mu$. In the present case in which  
$ \mu+P_{r}=0$, and $P_{\theta}=P_{\phi}=P_{\perp} $ the SEC are satisfied if
\begin {equation}
\mu+P_{\perp}\geq 0 \>\>\> \mbox {and}\>\>\>P_\perp\geq 0
\label {2-34}
\end {equation}
A large number of function $ H(r) $ which satisfy the matching conditions (\ref{2-4}) are given in Ref \cite{Ky2}.   

We shall present some Examples. The $ H(r)$ of these examples satisfy relations (\ref{2-4})

\section{Examples}

\subsection{Example 1}

The solution with the $ H(r)$  of Eq (\ref{3-1-1}), which is given below

\begin {equation}
H(r)=-2 M k(r/k+b(1-r/k)^2)
\label {3-1-1}
\end {equation}
For the above $ H(r)$ the expressions for $\mu$,  $P_{\perp}$ and $\mu+P_{\perp}$ of Eqs (\ref{2-21}), (\ref{2-23})  and (\ref{2-25}) in the variables 
$y$, $n$ and $u$  defined by the relations
\begin {equation}
y=r/k,\>\> n={2 M}/k\>\>\mbox{and}\>\>u=a/k
\label {3-1-2}
\end {equation}
are the following:
\begin {eqnarray}
\label {test}
&&\mu= -\frac {1} {2\pi k^2\left (-2 u^2 x^2 + 
       4\sin (1 - y) + \cos (2 - 2 y) - 3 \right)^2} n (b y^2 - 2 b y + \nonumber \\
  &&   (2 b (y - 1) + 1)\tan (1 - y) +(-2 b (y - 1) - 1)\sec (1 - y) + b + y )
\label {3-1-3}
\end {eqnarray}
\begin {eqnarray}
\label {test}
&& P_{\perp}=-\frac {1} {64\pi k^2\left (u^2 x^2 + \sin^2 (1 - y) - 
       2\sin (1 - y) + 1 \right)^2} n (\sec^3 (1 - y) (-4 b   \nonumber \\
   &&    \cos (1 - y) (-2 u^2 x^2 + 
          4\sin (1 - y) + \cos (2 - 2 y) - 3) - (2 b (y - 1) +  1) \nonumber \\
    &&      (2\sin (1 - y) (2 u^2 x^2 - 3\cos (2 - 2 y) + 1) + 
          8\cos (2 - 2 y))) + 8 (b (y - 1)^2   \nonumber \\
    &&      + y))
\label {3-1-4}
\end {eqnarray}
\begin {eqnarray}
\label {test}
&& \mu+P_{\perp}=  -\frac {1} {64\pi k^2\left (u^2 x^2 + (\sin (1 - y) - 2)\sin (1 - 
          y) + 1 \right)^2} n\sec^3 (1 - y)  \nonumber \\
 &&    (-4 b\cos (1 - y) (-2 u^2 x^2 + 
       4\sin (1 - y) + \cos (2 - 2 y) - 3 ) - (2 b (y - 1) \nonumber \\
   &&    + 1) (4 u^2 x^2\sin (1 - y) - 
       5\sin (3 - 3 y) + 3\sin (1 - y) + 12\cos (2 - 2 y) +  4 ) \nonumber \\
 &&  + 16 (b (y - 1)^2 + y )\cos^3 (1 - y)) 
\label {3-1-5}
\end {eqnarray}
Using numerical computer calculations we find that  in the  interior region we have
\begin {equation}
\mu\geq 0,\>\>\>\>\>\>P_{\perp}\geq 0,\>\>\>\>\>\> \mu+P_{\perp}\geq 0\>\>\>\>\>\> \mbox{for}\>\>\>b\leq 0
\label {3-1-6}
\end {equation}

Also with numerical computer calculations we find that there is no value of $b$ for which $\mu-P_{\perp}\geq 0$ in the whole interior region. Therefore the 
solution with the metric (\ref{2-1}) and the $ F(r) $ and $ H(r) $ of Eqs (\ref{2-2}) and (\ref{3-1-1}) with  $b\leq 0$ satisfies the WEC and the SEC but there is no 
value of $b$ for which the solution satisfies the DEC.

In addition the solution should satisfy the relation (\ref{2-20a}), which for the $F(r)$ and $H(r)$ of Eqs (\ref{2-2}) and (\ref{3-1-1}) respectively
 and the use of Eqs (\ref{3-1-2}) becomes
\begin {equation}
u^2> n (y-b (1-y))^2-(1-\sin (1-y))^2
\label {3-1-7}
\end {equation}

\begin{figure}[H]
\includegraphics[scale=0.80]{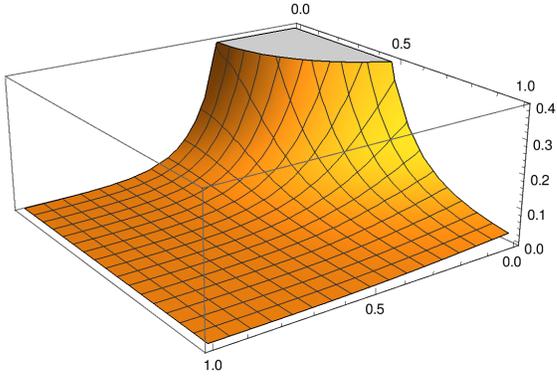}
\caption{ $ k^2 \mu $ of expression (\ref{3-1-3}) for $ (b=0, \>\>n=1,\>\>u=0.8\>) $ }
\label{f:4-1-1-1}
\end{figure}

\begin{figure}[H]
\includegraphics[scale=0.80]{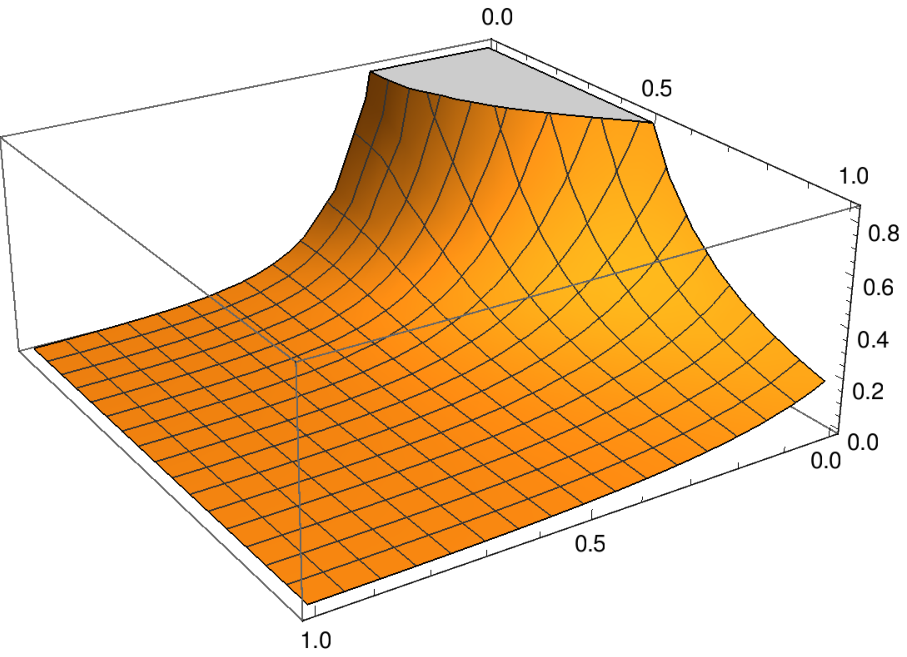}
\caption{ $ k^2 P_\perp $ of expression (\ref{3-1-4}) for $ (b=0,\>\>n=1,\>\>u=0.8\>) $ }
\label{f:4-1-1-2},
\end{figure}

\begin{figure}[H]
\includegraphics[scale=0.80]{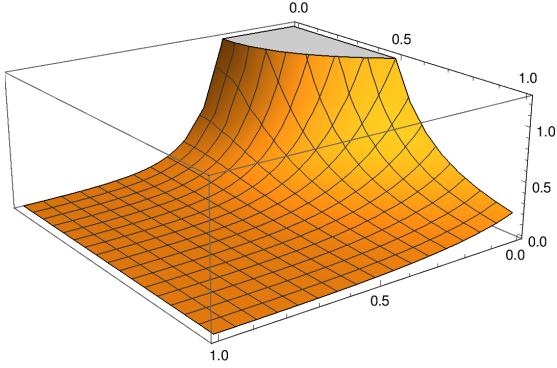}
\caption{ $ k^2(\mu+P_\perp) $ of expression (\ref{3-1-5})  for  $ (b=0,\>\>n=1,\>\>u=0.8\>) $ }
 \label{f:4-1-1-3}
\end{figure}

In Figures (\ref{f:4-1-1-1})-(\ref{f:4-1-1-3}) we present the graphs of $k^2\mu$, $k^2P_{\perp}$ and $k^2(\mu+P_{\perp})$ of Eqs 
(\ref{3-1-3}), (\ref{3-1-4}) and(\ref{3-1-5}) for $b=0$,  $n=1$ and $u=0.8$ for which relation (\ref{3-1-7}) is satisfied.
 With the help of a program \cite{Wo1} we can compute the minimum of
 these quantities for $0 \leq y \leq 1$ and $ 0 \leq  x^2 \leq 1$. We find the relations
.

$Mink^2\mu=3.38947*10^{-18}\>\>\> at\>\>\> y = 1, \>\>\>\mbox{and} \>\>\>x = 0.979865$

$Mink^2 P_\perp=0,\>\>\>at\>\>\> y = 1,\>\>\>\mbox{and}\>\>\> x = 9.15476*10^{-6} $

$Mink^2(\mu+P_{\perp})=0,\>\>\>at\>\>\> y = 1, \>\>\>\mbox{and}\>\>\> x = 4.10499*10^{-8}$

in agreement with relations (\ref{3-1-6}), which verify that the solution indeed satisfies the WEC and the SEC.

From the expression for $\mu$  of this example we find that $\mu$ for constant $x$  is monotonically decreasing for $ 0 \leq y \leq 1$ to the value $ \mu = 0$ at 
$y=1$. Also since $\mu+P_{r}=0$ the radial pressure $P_{r}$ is monotonically increasing to the value $P_{r}=0$ at $y=1$. The same thing happens to the examples
which follow. 

The line element of the solution is obtained from expression (\ref{2-1}) with $F(r)$ and $H(r)$ given by relations (\ref{2-2}) and (\ref{3-1-1}) respectively.
The line element for $b=0 $, which implies $ H(r)= -2 M r$, is the line element of the simplest solution of our class of solutions. 
Explicitly this line element is 

\begin {eqnarray}
\label {test}
&& d^2 s = -(1 - \frac {2 M r} {(1-sin(1-r/k))^2k^2 +  a^2 cos^2\theta}) d t^2  \nonumber \\
 &&        - \frac {4 a sin^2\theta  M r} {(1-sin(1-r/k))^2k^2 + a^2 cos^2\theta} dtd\phi  \nonumber \\
&&   + \frac {cos^2(1-r/k)((1-sin(1-r/k))^2k^2 + a^2 cos^2\theta)} {(1-sin(1-r/k))^2k^2 -2 M r  +  a^2} d r^2 \nonumber \\
 &&    + ((1-sin(1-r/k))^2k^2+a^2cos^2\theta)d \theta^2  + sin^2\theta ((1-sin(1-r/k))^2k^2  \nonumber \\
&&  +  a^2  + \frac { 2 a^2 sin^2\theta M r} {(1-sin(1-r/k))^2k^2 + a^2 cos^2\theta}) d \phi^2
\label {3-1-7'}
\end {eqnarray}
The above line element is obtained from the line element of the solution of Kerr if we multiply its coefficient of $d^2r$ by $ cos^2( 1-r/k) $ and replace its
$r^2$ by $ (1-sin(1-r/k))^2k^2 $.

\subsection {Example 2}

The solution with the $ H(r)$  of Eq (\ref{3-2-1}), which is given below

\begin {equation}
H(r)=-2 M k((1-b)e^{(r-k)/k}+br/k)
\label {3-2-1}
\end {equation}

For the above $H(r)$ the expressions for $\mu$, $P_{\perp}$ and $\mu+P_{\perp}$ of Eqs (\ref{2-21}), (\ref{2-23})  and (\ref{2-25}) in the variables 
$y$, $n$ and $u$ of Eqs (\ref{3-1-2}) become 

\begin {eqnarray}
\label {test}
&&\mu= -\frac {1} {2\pi k^2\left (-2 u^2 x^2 + 
       4\sin (1 - y) + \cos (2 - 2 y) - 
       3 \right)^2} n (b (-e^{y - 1})   \nonumber \\
 &&  + b y +  (b(-e^{y - 1}) + b + 
        e^{y - 1} )\tan (1 - y) + (b (e^{y - 1} - 1 ) - 
        e^{y - 1} )   \nonumber \\
  &&      \sec (1 - y) + e^{y - 1} )
\label {3-2-2}
\end {eqnarray}
\begin {eqnarray}
\label {test}
&&P_{\perp}= -\frac {1} {64\pi k^2\left (u^2 x^2 + \sin^2 (1 - y) - 
       2\sin (1 - y) + 1 \right)^2} n (\sec^3 (1 -  y)\nonumber \\
   &&    (2 (b - 1) e^{y - 1}\cos (1 - y) (-2 u^2 x^2 + 
          4\sin (1 - y) + \cos (2 - 2 y) - 3) + \nonumber \\
   &&       (b (e^{y - 1} - 1) - 
          e^{y - 1}) (2\sin (1 - y) (2 u^2 x^2 - 3\cos (2 - 2 y) + 1) + \nonumber \\
   &&          8\cos (2 - 2 y)) + 8 ((1 - b) e^{y - 1} + b y))
\label {3-2-3}
\end {eqnarray}
\begin {eqnarray}
\label {test}
&&\mu+P_{\perp}= -\frac {1} {64\pi k^2\left (u^2 x^2 + (\sin (1 - y) - 2)\sin (1 - 
          y) + 1 \right)^2} n\sec^3 (1 - y)  \nonumber \\
  &&  (2 (b - 1) e^{y - 1}\cos (1 - y) (-2 u^2 x^2 + 
       4\sin (1 - y) + \cos (2 - 2 y) - 3) - \nonumber \\
  &&     (b (1 - e^{y - 1}) + 
       e^{y - 1}) (4 u^2 x^2\sin (1 - y) - 5\sin (3 - 3 y) + 
       3\sin (1 - y) + \nonumber \\
  &&     12\cos (2 - 2 y) + 4) + 
    16 ((1 - b) e^{y - 1} + b y)\cos^3 (1 - y))
\label {3-2-4}
\end {eqnarray}
Using numerical computer calculations we find that in the interior region $0\leq y \leq 1$ and  $0 \leq x^2 \leq 1$ the following relations are satisfied:     
\begin {equation}
\mu\geq 0,\>\>\>P_{\perp} \geq 0,\>\>\>\mu+P_{\perp} \geq 0\>\>\>\mbox{for}\>\>\>b\geq 1
\label {3-2-5}
\end {equation}
Therefore the solution for $b\geq 1$ satisfies the WEC and the SEC. Also using numerical computer calculations we find that there is no value of $b$  for which 
$\mu-P_{\perp} \geq 0$ in the interior region. Therefore the solution does not satisfies the DEC.

According to our previous arguments the relation (\ref{2-20a}) should hold. For $F(r)$ and $H(r)$ given by Eqs (\ref{2-2}) and (\ref{3-2-1}) respectively
and the use of Eqs (\ref{3-1-2})
this relation becomes 
\begin {equation}
u^2 > n \left((1-b) e^{y-1}+b y\right)-(1- \sin(1-y) )^2
\label {3-2-6}
\end {equation}

\begin{figure}[H]
\includegraphics[scale=0.80]{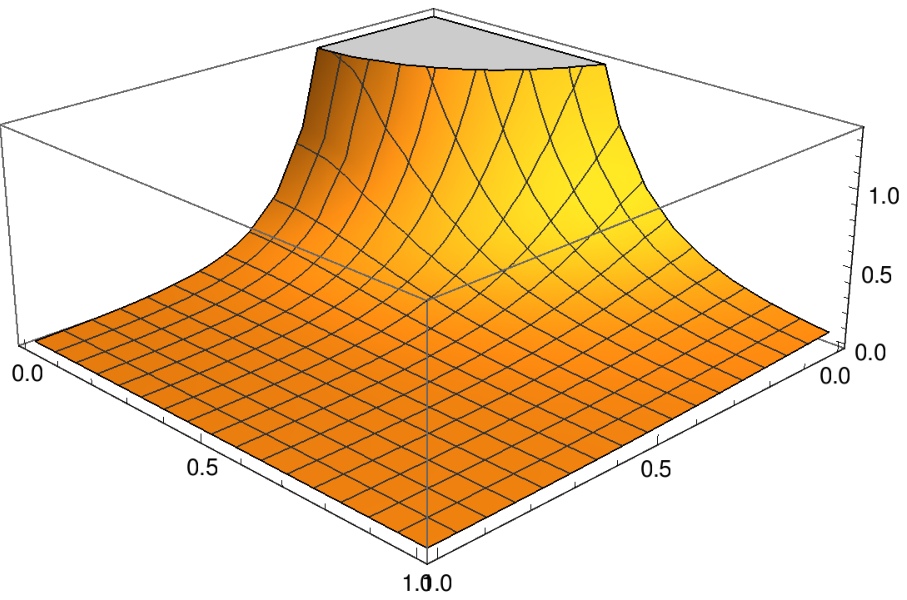}
\caption{ $ k^2 \mu $ of expression (\ref{3-2-2}) for $ (b=2,\>\>n=1,\>\>u=0.8\>) $ }
\label{f:3-2-1}
\end{figure}

\begin{figure}[H]
\includegraphics[scale=0.80]{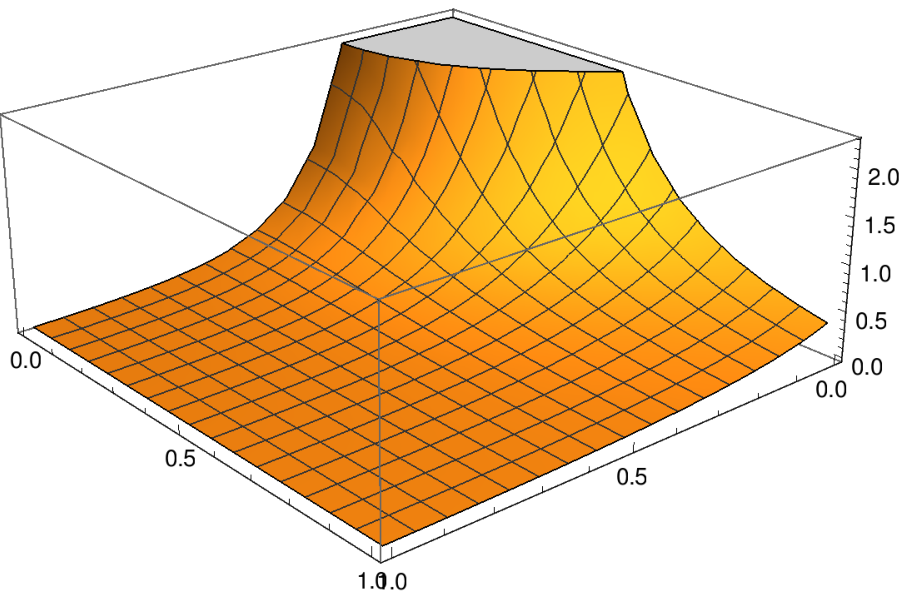}
\caption{ $ k^2 P_\perp $ of expression (\ref{3-2-3}) for $ (b=2,\>\>n=1,\>\>u=0.8\>) $ }
\label{f:3-2-2}
\end{figure}

\begin{figure}[H]
\includegraphics[scale=0.80]{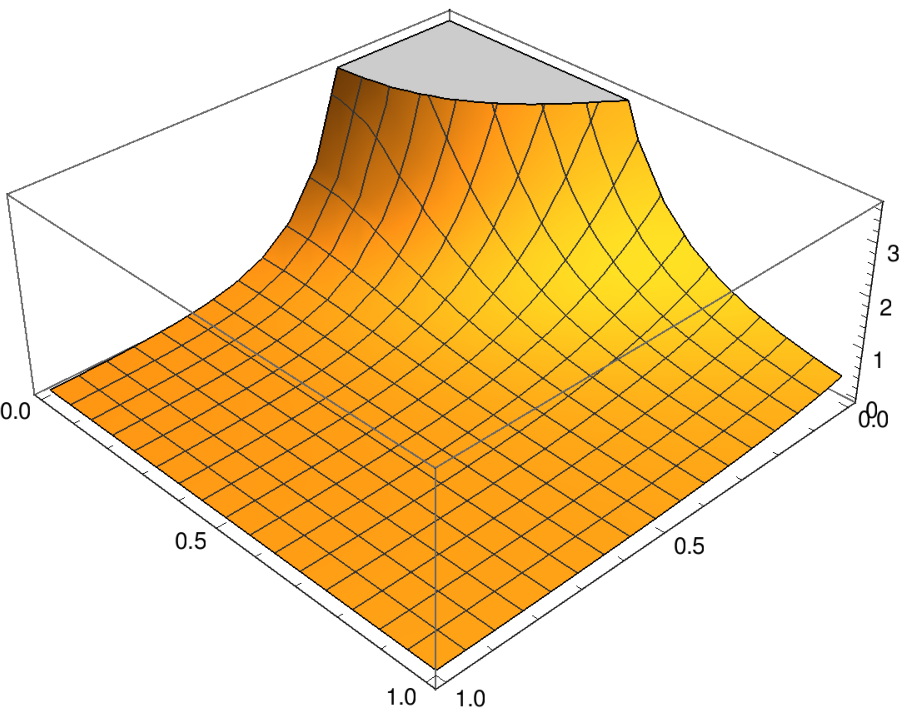}
\caption{ $ k^2(\mu+P_\perp) $ of expression (\ref{3-2-4})  for  $ (b=2,\>\>n=1,\>\>u=0.8\>) $ }
 \label{f:3-2-3}
\end{figure}

The graphs of $k^2\mu$, $k^2P_{\perp}$ and $k^2(\mu+P_{\perp})$ are presented in Figures  (\ref{f:3-2-1}), (\ref{f:3-2-2}) and (\ref{f:3-2-3})  for
$b=2$, $n=1$ and $u=0.8$, which satisfy relation (\ref{3-2-6}). With the help of a program \cite{Wo1} we can compute the minimum of these quantities for
$0 \leq y \leq 1$ and $ 0 \leq  x^2 \leq 1$.
We find the relations 

$Mink^2\mu=0\>\>\> at\>\>\> y = 1 \>\>\>and \>\>\>x =3.38904*10^{-9} $

$Mink^2P_{\perp}=0.0121307,\>\>\>at\>\>\> y = 1,\>\>\>and\>\>\> x = 1 $

$Mink^2(\mu+P_{\perp})=0.0121307,\>\>\>at\>\>\> y = 1, \>\>\>and\>\>\> x =1$

in agreement with relations (\ref{3-2-5}), which verify that the solution indeed satisfies the WEC and the SEC. 

The line element of the solution is that of Eq (\ref{2-1}) where $ F(r)$ and $H(r)$ are given by Eqs (\ref{2-2}) and (\ref{3-2-1}) respectively.

\subsection{Example 3}

The solution with the $ H(r)$  of Eq (\ref{3-3-1}), which is given below

\begin {equation}
H(r)=-\frac{2 M}{b} k^{1-b} \left((b-1) k^b+r^b\right)
\label {3-3-1}
\end {equation}

For the above $H(r)$ the expressions for $\mu$, $P_{\perp}$ and $\mu+P_{\perp}$ of Eqs (\ref{2-21}), (\ref{2-23})  and (\ref{2-25}) in the variables 
$y$, $n$ and $u$ of Eqs (\ref{3-1-2}) become 

\begin {equation}
\mu=-\frac {n(yt (y^b + b - 1 ) + b y^b\tan (1 - y) - 
     b y^b\sec (1 - y) )} {2\pi b k^2 y (-2 u^2 x^2 + 
      4\sin (1 - y) + \cos (2 - 2 y) - 3 )^2}
\label {3-3-2}
\end {equation}

\begin {eqnarray}
\label {test}
&& P_{\perp}= \frac {1} {32\pi k^2\left (u^2 x^2 + (\sin (1 - y) - 2)\sin (1 - 
          y) + 1 \right)^2} n (y^{b - 2}\sec^2 (1 - y) \nonumber \\
  &&     (-2 b u^2 x^2 + 
       4 (b - 1)\sin (1 - y) + \cos (2 - 2 y) (b - 3 y\tan (1 - y) + \nonumber \\
  &&       4 y\sec (1 - y) - 1) - 3 b + 2 u^2 x^2 y\tan (1 - y) + 
       2 u^2 x^2 + y\tan (1 - y) +  3) \nonumber \\
    &&   - \frac {4 (y^b + b - 1)} {b}) 
\label {3-3-3}
\end {eqnarray}

\begin {eqnarray}
\label {test}
&&\mu+P_{\perp}= -\frac {1} {64\pi k^2\left (u^2 x^2 + (\sin (1 - y) - 2)\sin (1 - 
          y) + 1 \right)^2} n\sec^3 (1 - y)  \nonumber \\
 &&   (-2 (b - 1) y^{b - 2}\cos (1 - y) (-2 u^2 x^2 + 
       4\sin (1 - y) + \cos (2 - 2 y) - 3) -  \nonumber \\
  &&  y^{b - 1} (4 u^2 x^2\sin (1 - y) - 5\sin (3 - 3 y) + 
       3\sin (1 - y) + 12\cos (2 - 2 y) + 4)  \nonumber \\
  &&     + \frac {16 (y^b + b - 1)\cos^3 (1 - 
         y)} {b}) 
\label {3-3-4}
\end {eqnarray}

Using numerical computer calculations we find that in the interior region $0\leq y \leq 1$ and  $0 \leq x^2 \leq 1$ the following relations are satisfied:     
\begin {equation}
\mu\geq 0,\>\>\>P_{\perp} \geq 0,\>\>\>\mu+P_{\perp} \geq 0\>\>\>\mbox{for}\>\>\>0< b \leq 1
\label {3-3-5}
\end {equation}
Therefore the solution for $ 0< b \leq 1$ satisfies the WEC and the SEC. Also using numerical computer calculations we find that there is no value of $b$  for which 
$\mu-P_{\perp} \geq 0$ in the interior region. Therefore the solution does not satisfies the DEC.

Also the relation (\ref{2-20a}) should hold. For $F(r)$ and $H(r)$ given by Eqs (\ref{2-2}) and (\ref{3-3-1}) respectively and the use of Eqs (\ref{3-1-2}) 
this relation becomes 
\begin {equation}
u^2 > \frac{n }{b}\left(y^b+b-1\right)-(1-\sin (1-y))^2
\label {3-3-6}
\end {equation}

\begin{figure}[H]
\includegraphics[scale=0.80]{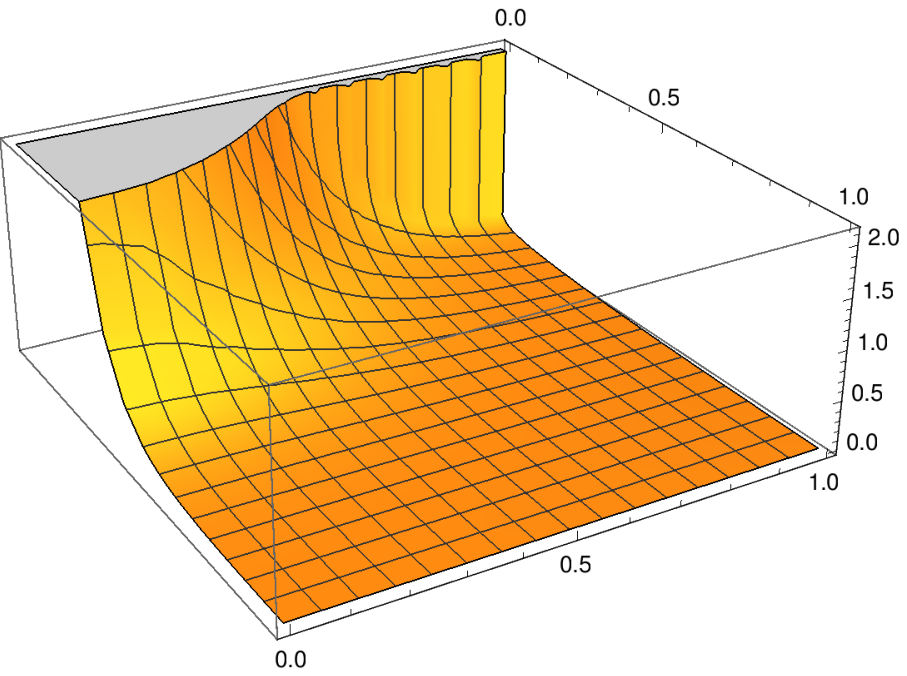}
\caption{ $ k^2 \mu $ of expression (\ref{3-3-2}) for $ (b=0.5,\>\>n=1,\>\>u=0.8\>) $ }
\label{f:3-3-1}
\end{figure}

\begin{figure}[H]
\includegraphics[scale=0.80]{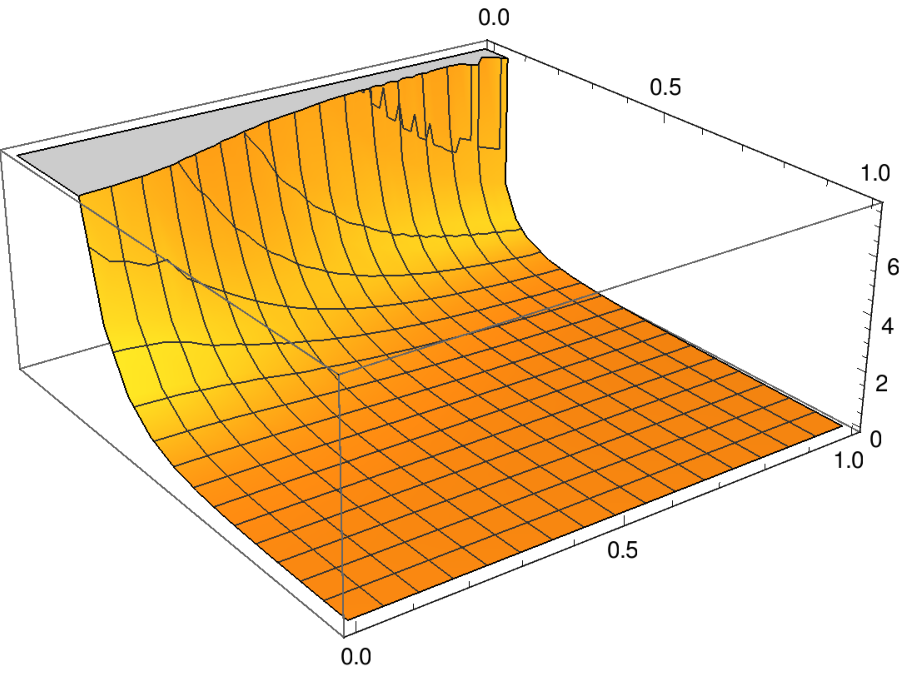}
\caption{ $ k^2 P_\perp $ of expression (\ref{3-3-3}) for $ (b=0.5,\>\>n=1,\>\>u=0.8\>) $ }
\label{f:3-3-2}
\end{figure}

\begin{figure}[H]
\includegraphics[scale=0.80]{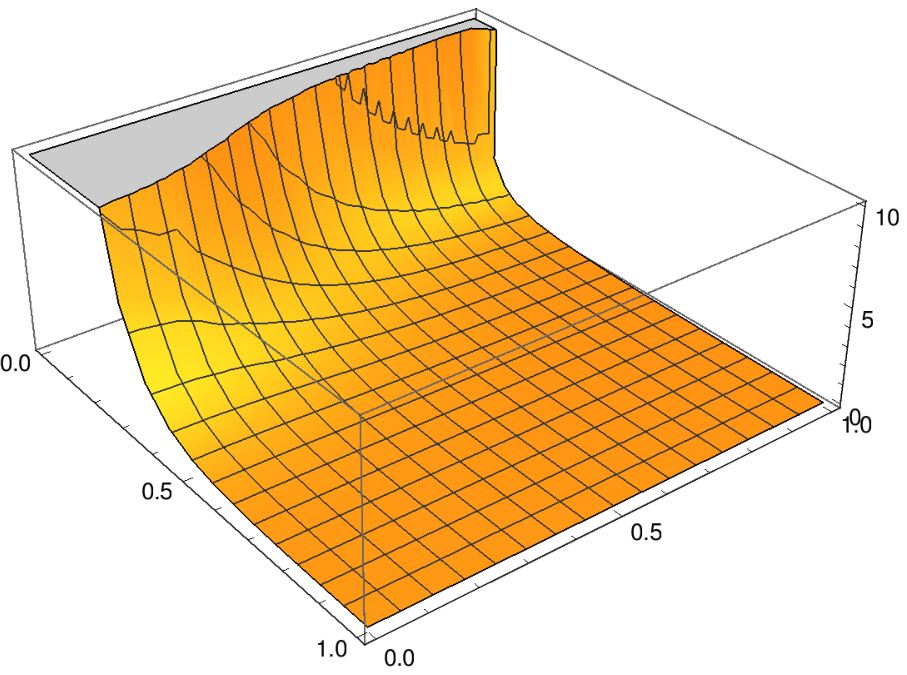}
\caption{ $ k^2(\mu+P_\perp) $ of expression (\ref{3-3-4})  for  $ (b=0.5,\>\>n=1,\>\>u=0.8\>) $ }
 \label{f:3-3-3}
\end{figure}

The graphs of $k^2\mu$, $k^2P_{\perp}$ and $k^2(\mu+P_{\perp})$ are presented in Figures  (\ref{f:3-3-1}), (\ref{f:3-3-2}) and (\ref{f:3-3-3})  for
$b=0.5$, $n=1$ and $u=0.8$, which satisfy relation (\ref{3-3-6}). With the help of a program \cite{Wo1} we can compute the minimum of these quantities 
for $0 \leq y \leq 1$ and $ 0 \leq  x^2 \leq 1$.
We find the relations 

$Mink^2\mu=0\>\>\> at\>\>\> y = 1 \>\>\>and \>\>\>x =4.66932*10^{-8} $

$Mink^2P_{\perp}=0.00606536,\>\>\>at\>\>\> y = 1,\>\>\>and\>\>\> x = 1 $

$Mink^2(\mu+P_{\perp})=0.00606536,\>\>\>at\>\>\> y = 1, \>\>\>and\>\>\> x =1$

in agreement with relations (\ref{3-3-5}), which verify that the solution indeed satisfies the WEC and the SEC.

The line element of the solution is that of Eq (\ref{2-1}) where $ F(r)$ and $H(r)$ are given by Eqs (\ref{2-2}) and (\ref{3-3-1}) respectively.

\subsection{Example 4}

The solution with the $ H(r)$ of Eq (\ref{3-4-1}), which is given below

\begin{equation}
H(r)=-2 M \left(b k e^{\frac{r-k}{b k}}-b k+k\right)
\label {3-4-1}
\end {equation}

For the above $H(r)$ the expressions for $\mu$, $P_{\perp}$ and $\mu+P_{\perp}$ of Eqs (\ref{2-21}), (\ref{2-23})  and (\ref{2-25}) in the variables 
$y$, $n$ and $u$ of Eqs (\ref{3-1-2}) become 
\begin {equation}
\mu=\frac {n\left (b\left (-e^{\frac {y - 1} {b}} \right) - 
     e^{\frac {y - 1} {b}}\tan (1 - y) + 
     e^{\frac {y - 1} {b}}\sec (1 - y) + b - 
     1 \right)} {8\pi k^2\left (u^2 x^2 + \sin^2 (1 - y) - 
      2\sin (1 - y) + 1 \right)^2}
\label {3-4-2}
\end {equation}
\begin {eqnarray}
\label {test}
&& P_{\perp}=\frac {1} {8\pi b k^2 (-2 u^2 x^2 + 
       4\sin (1 - y) + \cos (2 - 2 y) - 3)^2} n (\sec^2 (1 - y) \nonumber \\
   &&    (b (2 u^2 x^2 + 1)\tan (1 - y) + \cos (2 - 2 y) (1 - 3 b\tan (1 - y)) - 2 u^2 x^2 - 3)  \nonumber \\
    &&   + 4 b (b (-e^{\frac {y - 1} {b}}) + b - 1) + 
    4 b\cos (2 - 2 y)\sec^3 (1 - y) + 
    4\tan (1 - y) \nonumber \\
 &&   \sec (1 - y)) 
\label {3-4-3}
\end {eqnarray}
\begin {eqnarray}
 \label {test}
 && \mu+P_{\perp}= \frac {1} {16\pi b k^2 (-2 u^2 x^2 + 
           4\sin (1 - y) + \cos (2 - 2 y) - 
           3)^2} n (e^{\frac {y - 1} {b}}  \nonumber \\
      &&     \sec^2 (1 - y) (b (4 u^2 x^2 + 3)\tan (1 - y) - 4 u^2 x^2 + 
           2\cos (2 - 2 y) - 6) + 16 b   \nonumber \\
      &&     (b (-e^{\frac {y - 1} {b} } +
   b - 1) + 8 e^{\frac {y - 1} {b}}\tan (1 - y)\sec (1 - y) + 
       b e^{\frac {y - 1} {b}}\sec^3 (1 - y) (-5  \nonumber \\
 &&     \sin (3 - 3 y) + 
    12\cos (2 - 2 y) + 4)) 
 \label {3-4-4}
\end {eqnarray}

With numerical computer calculations we find that in the interior region $0\leq y \leq 1$ and  $0 \leq x^2 \leq 1$ the following relations are satisfied:     
\begin {equation}
\mu\geq 0,\>\>\>P_{\perp} \geq 0,\>\>\>\mu+P_{\perp} \geq 0\>\>\>\mbox{for}\>\>\> b < 0
\label {3-4-5}
\end {equation}
Therefore the solution for $ b < 0$ satisfies the WEC and the SEC. Also with numerical computer calculations we find that there is no value of $b$  for which 
$\mu-P_{\perp} \geq 0$ in the interior region. Therefore the solution does not satisfies the DEC.

The relation (\ref{2-20a}), which should hold, becomes for $F(r)$ and $H(r)$ given by Eqs (\ref{2-2}) and (\ref{3-4-1}) respectively and the use of 
Eqs (\ref{3-1-2})  
\begin {equation}
u^2 > n \left(b e^{\frac{y-1}{b}}-b+1\right)-((y-1) \sin +1)^2
\label {3-4-6}
\end {equation}

\begin{figure}[H]
\includegraphics[scale=0.80]{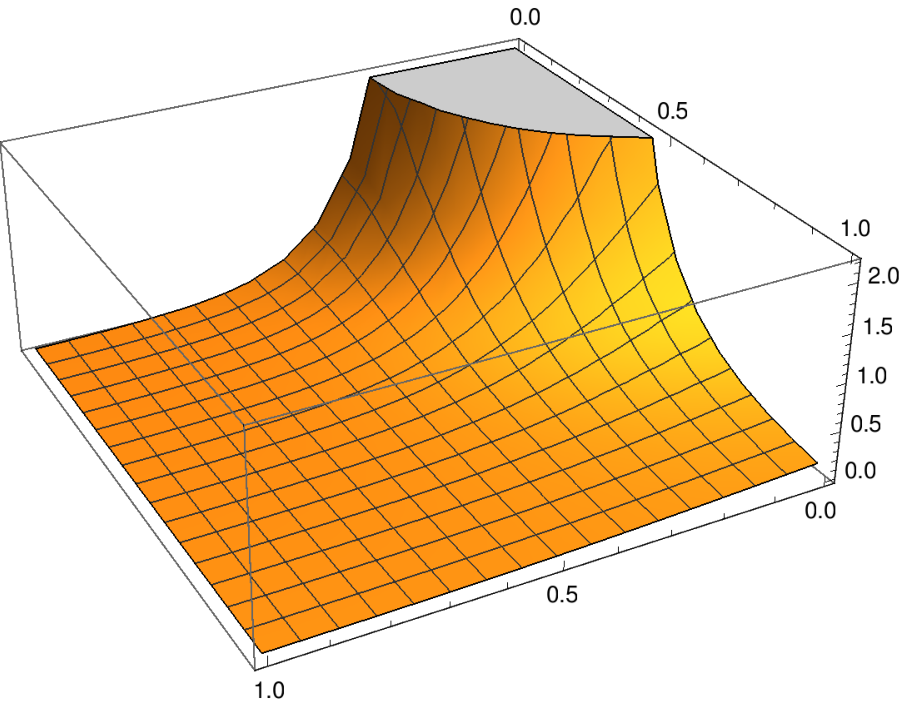}
\caption{ $ k^2 \mu $ of expression (\ref{3-4-2}) for $ (b=-1,\>\>n=1,\>\>u=0.8\>) $ }
\label{f:3-4-1}
\end{figure}

\begin{figure}[H]
\includegraphics[scale=0.80]{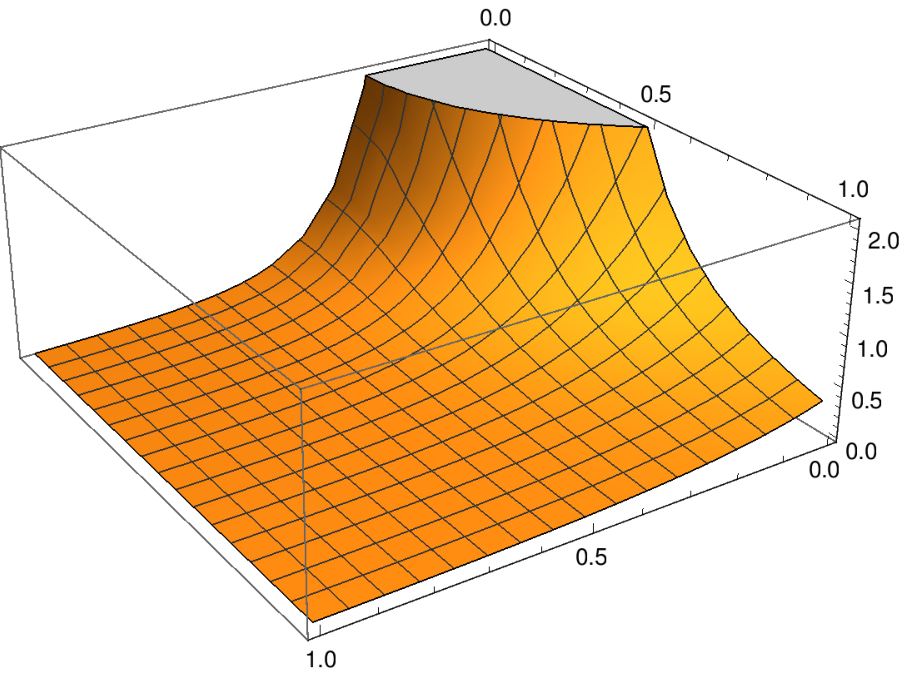}
\caption{ $ k^2 P_\perp $ of expression (\ref{3-4-3}) for $ (b=-1,\>\>n=1,\>\>u=0.8\>) $ }
\label{f:3-4-2}
\end{figure}

\begin{figure}[H]
\includegraphics[scale=0.80]{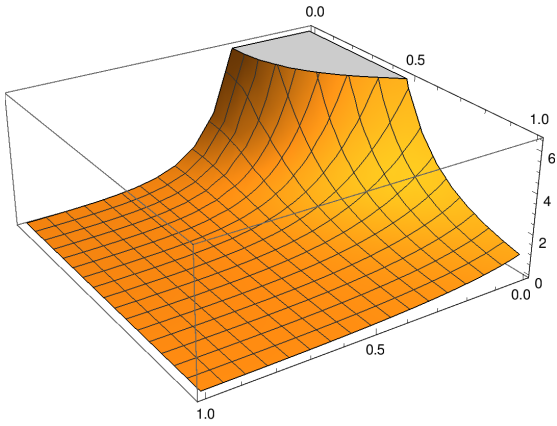}
\caption{ $ k^2(\mu+P_\perp) $ of expression (\ref{3-4-4})  for  $ (b=-1,\>\>n=1,\>\>u=0.8\>) $ }
 \label{f:3-4-3}
\end{figure}

The graphs of $k^2\mu$, $k^2P_{\perp}$ and $k^2(\mu+P_{\perp})$ are presented in Figures  (\ref{f:3-4-1}), (\ref{f:3-4-2}) and (\ref{f:3-4-3})  for
$b=-1$, $n=1$ and $u=0.8$, which satisfy the above relation. With the help of a program \cite{Wo1} we can compute the minimum of these quantities for
$0 \leq y \leq 1$ and $ 0 \leq  x^2 \leq 1$. We find the relations 

$Mink^2\mu=0\>\>\> at\>\>\> y = 1 \>\>\>and \>\>\>x =2.36449*10^{-8} $

$Mink^2P_{\perp}=0.0121307,\>\>\>at\>\>\> y = 1,\>\>\>and\>\>\> x = 1 $

$Mink^2(\mu+P_{\perp})=0.0121307,\>\>\>at\>\>\> y = 1, \>\>\>and\>\>\> x =1$

in agreement with relations (\ref{3-4-5}), which verify that the solution indeed satisfies the WEC and the SEC.             

The line element of the solution is that of Eq (\ref{2-1}) where $ F(r)$ and $H(r)$ are given by Eqs (\ref{2-2}) and (\ref{3-4-1}) respectively.

\subsection{Example 5}

The solution with the $ H(r)$  of Eq (\ref{3-5-1}), which is given below

\begin{equation}
H(r)=-\frac{2 k M (b r-k)}{(b-2) k+r}
\label {3-5-1}
\end {equation}

For the above $H(r)$ the expressions for $\mu$, $P_{\perp}$ and $\mu+P_{\perp}$ of Eqs (\ref{2-21}), (\ref{2-23})  and (\ref{2-25}) in the variables 
$y$, $n$ and $u$ of Eqs (\ref{3-1-2}) become 
\begin {eqnarray}
\label {test}
&&\mu= -\frac {n} {2\pi k^2 (b + y - 2)^2(-2 u^2 x^2 + 
       4\sin (1 - y) + \cos (2 - 2 y) - 3 )^2}  (b^2 y + \nonumber \\
 &&   b y^2 - 2 b y + (b - 1)^2\tan (1 - y) + (b - 1)^2 (-\sec (1 - y)) - b - y +  2 ) 
\label {3-5-2}
\end {eqnarray}
\begin {eqnarray}
\label {test}
&&P_{\perp}= \frac {n} {64\pi k^2 (b + y - 
        2)^3\left (u^2 x^2 + (\sin (1 - y) - 2)\sin (1 - y) + 
       1 \right)^2} ((b - 1)^2  \nonumber \\
   &&    \sec^3 (1 - y) ((b + y - 
          2) (2\sin (1 - y) (2 u^2 x^2 - 3\cos (2 - 2 y) + 1) + \nonumber \\
     &&     8\cos (2 - 2 y)) - 4\cos (1 - y) (-2 u^2 x^2 + 4\sin (1 - y) + \cos (2 - 2 y) -  3)) +  \nonumber \\
      &&    8 (b + y - 2)^2 (1 - b y)) 
\label {3-5-3}
\end {eqnarray}
\begin {eqnarray}
\label {test}
 && \mu+P_{\perp}= \frac {n} {64\pi k^2 (b + y - 
            2)^3\left (u^2 x^2 + (\sin (1 - y) - 2)\sin (1 - y) + 1 \right)^2}  \nonumber \\
      &&     \sec^3 (1 - y) -  4 (b - 1)^2\cos (1 - y) (-2 u^2 x^2 + 4\sin (1 - y) + \cos (2 - 2 y) \nonumber \\
   &&    - 3) + (b - 1)^2 (b + y - 
       2) (4 u^2 x^2\sin (1 - y) - 5\sin (3 - 3 y) + 3\sin (1 - y)  \nonumber \\
   &&   +12\cos (2 - 2 y) + 4) -  16 (b + y - 2)^2 (b y - 1)\cos^3 (1 - y)) 
\label {3-5-4}
\end {eqnarray}

With numerical computer calculations we find that in the interior region $0\leq y \leq 1$ and  $0 \leq x^2 \leq 1$ the following relations are satisfied:     
\begin {equation}
\mu\geq 0,\>\>\>P_{\perp} \geq 0,\>\>\>\mu+P_{\perp} \geq 0\>\>\>\mbox{for}\>\>\> b > 2
\label {3-5-5}
\end {equation}
which imply that the solution for $ b > 2$ satisfies the WEC and the SEC. Also with numerical computer calculations we find that there is no value of $b$  for which 
$\mu-P_{\perp} \geq 0$ in the interior region. Therefore the solution does not satisfies the DEC.

The relation (\ref{2-20a}), which should hold, becomes for $F(r)$ and $H(r)$ given by Eqs (\ref{2-2}) and (\ref{3-5-1}) respectively and the use of
 Eqs (\ref{3-1-2}) 
\begin {equation}
u^2 >-\frac{n-b n y}{b+y-2}-(1-\sin (1-y))^2
\label {3-5-6}
\end {equation}

\begin{figure}[H]
\includegraphics[scale=0.80]{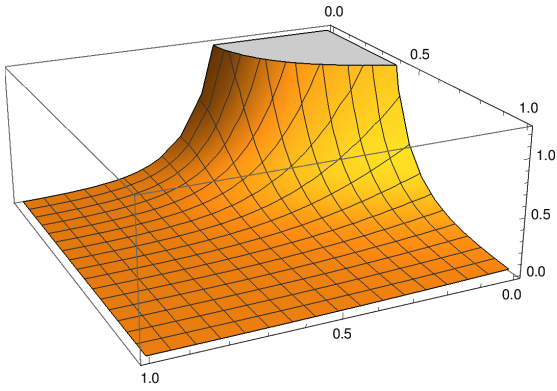}
\caption{ $ k^2 \mu $ of expression (\ref{3-5-2}) for $ (b=3,\>\>n=1,\>\>u=1.2\>) $ }
\label{f:3-5-1}
\end{figure}

\begin{figure}[H]
\includegraphics[scale=0.80]{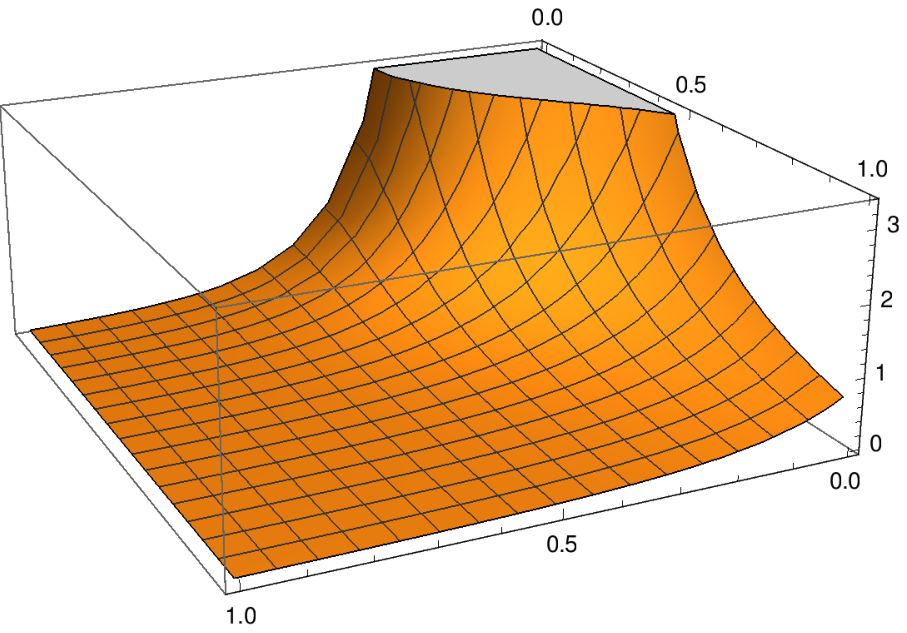}
\caption{ $ k^2 P_\perp $ of expression (\ref{3-5-3}) for $ (b=3,\>\>n=1,\>\>u=1.2\>) $ }
\label{f:3-5-2}
\end{figure}

\begin{figure}[H]
\includegraphics[scale=0.80]{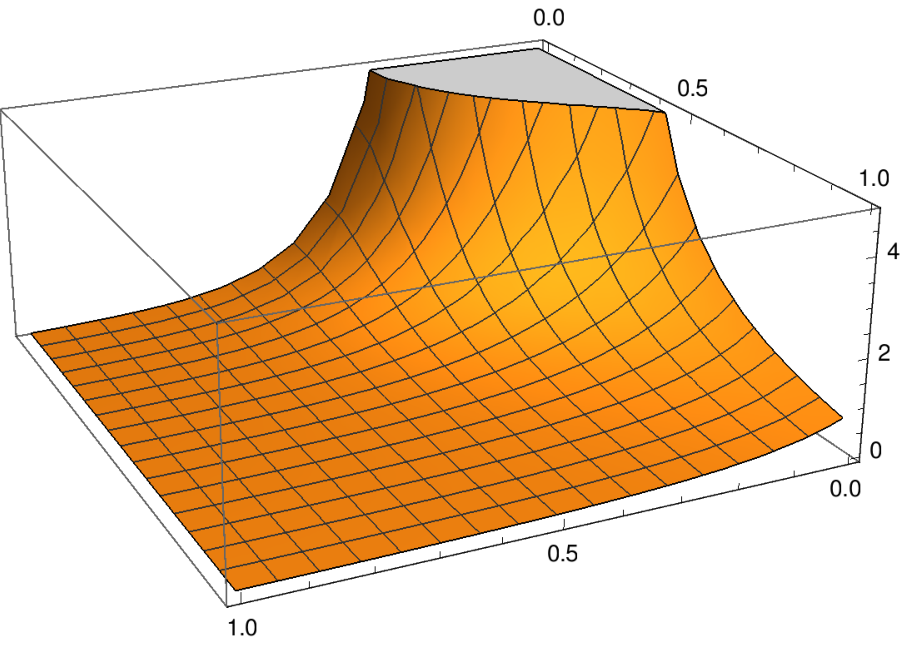}
\caption{ $ k^2(\mu+P_\perp) $ of expression (\ref{3-5-4})  for  $ (b=3,\>\>n=1,\>\>u=1.2\>) $ }
 \label{f:3-5-3}
\end{figure}

The graphs of $k^2\mu$, $k^2P_{\perp}$ and $k^2(\mu+P_{\perp})$ are presented in Figures  (\ref{f:3-5-1}), (\ref{f:3-5-2}) and (\ref{f:3-5-3})  for
$b=3$, $n=1$ and $u=1.2$, which satisfy the above relation. With the help of a program \cite{Wo1} we can compute the minimum of these quantities for
$0 \leq y \leq 1$ and $ 0 \leq  x^2 \leq 1$. We find the relations

$Mink^2\mu=0\>\>\> at\>\>\> y = 1 \>\>\>and \>\>\>x =8.62525*10^{-8} $

$Mink^2P_{\perp}=0.00815343,\>\>\>at\>\>\> y = 1,\>\>\>and\>\>\> x = 1 $

$Mink^2(\mu+P_{\perp})=0.00815343,\>\>\>at\>\>\> y = 1, \>\>\>and\>\>\> x =1$

in agreement with relations (\ref{3-5-5}), which verify that the solution indeed satisfies the WEC and the SEC.             

The line element of the solution is that of Eq (\ref{2-1}) where $ F(r)$ and $H(r)$ are given by Eqs (\ref{2-2}) and (\ref{3-5-1}) respectively.

\end{document}